\documentstyle[11pt]{article}

\begin{document}

\setlength{\baselineskip}{0.8 cm}

\begin{center}
{\Large \bf Dirac Monopoles and the Angular Momentum of the Electromagnetic
Field\footnote{Work partially
supported by CNPq and FAPESP}}
\end{center}

\begin{center}
{\large \bf M.C. Nemes$^{(1)}$ and
Saulo C.S. Silva$^{(2)}$}
\end{center}

\begin{center}
$^{(1)}$Instituto de Ci\^encias Exatas, Universidade Federal de Minas
Gerais\\
CP702, 30000,
Belo Horizonte, MG, Brasil.
\end{center}

\begin{center}
$^{(2)}$Instituto de F\'{\i}sica, Universidade de S\~ao Paulo\\
CP20516, 01498, S\~ao Paulo, SP, Brasil.
\end{center}

\begin{abstract}
We present a possible solution for the long standing problem of the
incompatibility of Dirac's charge quantization condition with integer values
for the angular momentum of the electromagnetic field.
\vspace{0.5 cm}

\end{abstract}

Dirac's charge quantization condition$^{[1]}$ can be obtained semiclassically
 if one
considers the scattering of an electric charge $e$ in the magnetic monopole's
field$^{[2]}$. The variation of the charge's angular momentum is given as

\begin{equation}
\Delta\vec{L}_{e}=\frac{eg}{2\pi}\hat{e}_{z}
\end{equation}

\noindent
where $\hat{e}_{z}$ gives the direction of the charge's trajectory. The
conservation of the total angular momentum of the system is guaranteed by
an equal and opposite variation on the field's angular momentum.
In fact, the initial and final angular momentum of the electromagnetic field
is given by$^{[3,4]}$

\begin{equation}
\vec{L}_{em}=-\frac{eg}{4\pi}\frac{z}{|z|}\hat{e}_{z}
\end{equation}

\noindent
so that the variation in this quantity during the process is

\begin{equation}
\Delta\vec{L}_{em}=-\frac{eg}{2\pi}\hat{e}_{z}
\end{equation}

A problem arises when one confronts $(2)$ with Dirac's charge quantization
condition

\begin{equation}
\frac{eg}{2\pi}=n
\end{equation}

It is easy to verify that one is immediately led to half integer values for
the field angular momentum.

The purpose of this brief report is to show that no such difficulties arise
when we take into account the angular momentum of the field {\it inside} the
string which is attached to the monopole$^{[5]}$.

Inside the string the magnetic field is given by

\begin{equation}
\vec{H}=\frac{g}{A}\hat{e}_{x}
\end{equation}

\noindent
where we assume the monopole to be at the origin, the string on
the negative valued side of the $x$-axis and A represents the string's cross
section. In the asymptotic region, i.e., when the charge is sufficiently
distant from the monopole, its electric field is given by ($v\ll1$)

\begin{equation}
\vec{E}=\frac{e}{4\pi}\frac{\vec{r}-z\hat{e}_{z}}{|\vec{r}-z\hat{e}_{z}|^{3}}
\end{equation}

\noindent
(the charge is assumed to be moving in the $x$-$z$ plane).

On the string we have

\begin{equation}
\vec{r}=x\hat{e}_{x}
\end{equation}

\noindent
and

\begin{equation}
\vec{E}=\frac{e}{4\pi}\frac{x\hat{e}_{x}-z\hat{e}_{z}}{(x^{2}+z^{2})^{\frac
{3}{2}}}
\end{equation}

We are now in a position to calculate the field's angular momentum in the
string:

\[\vec{L}_{string}=\int_{string}\vec{r}\times(\vec{E}\times\vec{H})dV]

\begin{equation}
=\frac{eg}{4\pi A}\hat{e}_{z}\int_{string}\frac{xz}{(x^{2}+z^{2})
^{\frac{3}{2}}}dV=\frac{egz}{4\pi}\int_{-\infty}^{0}\frac{xdx}{(x^{2}+z^{2})
^{\frac{3}{2}}}\hat{e}_{z}
\end{equation}

\[=\frac{eg}{4\pi}\frac{z}{|z|}\hat{e}_{z}]

The field's angular momentum will then be the sum (see $(2)$ and $(9)$)

\begin{equation}
\vec{L}_{field}=\vec{L}_{em}+\vec{L}_{string}=0
\end{equation}

The above result is not at all incompatible with Dirac's charge quantization
condition and does not require half integer values for $\vec{L}_{field}$.

A natural question now arises as to how the variation of the charge's angular
momentum is compensated. In order to answer this question one has to take the
angular momentum of the string into consideration. The string can be thought
of as a semi-infinite solenoid, or also as several magnetic dipoles aligned
$^{[6]}$. During the scattering process, the string tends to align with the
charge's magnetic field, and it acquires an angular momentum in $z$ direction,
opposed to $\Delta\vec{L}_{e}$, thus conserving the total angular momentum.

We believe to have presented a possible solution to a long standing problem
in the literature$^{[7]}$.

\end{document}